# On Anomalies in Annotation Systems

Matthias R. Brust and Steffen Rothkugel
*Faculty of Science, Technology and Communication (FSTC)*
*University of Luxembourg*
*Luxembourg*
*{matthias.brust, steffen.rothkugel}@uni.lu*

## Abstract

*Today's computer-based annotation systems implement a wide range of functionalities that often go beyond those available in traditional paper-and-pencil annotations. Conceptually, annotation systems are based on thoroughly investigated psycho-sociological and pedagogical learning theories. They offer a huge diversity of annotation types that can be placed in textual as well as in multimedia format. Additionally, annotations can be published or shared with a group of interested parties via well-organized repositories. Although highly sophisticated annotation systems exist both conceptually as well as technologically, we still observe that their acceptance is somewhat limited. In this paper, we argue that nowadays annotation systems suffer from several fundamental problems that are inherent in the traditional paper-and-pencil annotation paradigm. As a solution, we propose to shift the annotation paradigm for the implementation of annotation system.*

## 1. Introduction

The most frequently employed learning technique applied is that of *annotating*. Adding some keywords to a book's sidebar while reading or reviewing is one of the most common practices. An annotation is a textual and semiotic manifestation of a thought or idea that is often closely related to where the annotation is placed, i.e. its context. Hence, annotations can be considered as a constructive cognitive effort and represent key elements of constructive-based learning theories [1]. In terms of constructivism, annotations are essential learning mechanisms with many different purposes like: third-party comments, information filtering, semantic labeling, etc. [2]. Practically, annotations augment existing knowledge by additional information that is also visible and reusable for other persons.

The question may arise why annotations are so frequently used. Studies have demonstrated benefits in learning more because of the ability of *accessing* proper annotations than because of the activity of *writing them down*. Furthermore, annotations—although small in appearance—are powerful helpers. Often people need only keywords in a context to trigger sophisticated mental processes that remind them in the entirety of thoughts present when writing the annotation down. In that sense, annotations appear as anchor points with execution functionality for our mental process. Others would claim annotations have a mnemonic nature. An example of the potential and implications of an annotation is known as Fermat's last theorem [1].

In 1637, the mathematician Pierre de Fermat wrote a comment in the margin of the book "Arithmetica". This annotation was of the length of two normal-sized sentences, but it took more than 350 years and thousands of person-hours to prove the affirmation stated there. Interestingly, the comment became known when in 1670 Fermat's son published a new edition of that book augmented with comments done by his father [3]. This shows the immense potential of annotations to keep complex thoughts in as few words and lines as possible, as well as, illustrating the highly constructive character annotations can have.

Nowadays, these circumstances have resulted in the invention of annotation systems. Annotation systems are tools, aimed at bringing the comfort of paper annotations to the computer. A broad range of annotations systems exists and their functionalities can differ significantly.

Annotation systems usually provide various types of annotations to express their functional intention. For instance, there might be *question*, *advice*, *example*, and *correction* annotation types. The implementation of these types reflects the underlying conception and learning theory. A more elaborate concept that can be found in a few systems is that of replying to an already existing annotation. A crucial point is the location where annotations are placed on a digital document. Nowadays, most systems are able to perform on-page, in-line, and in-place annotations corresponding to annotations made on paper. Several annotation systems aim at supporting collaborative work. As such, they offer a common repository to share annotations across group members. In some systems, the annotations can even be declared as public, i.e. accessible to anyone. Many systems support advanced searching functionality. The user can easily search for annotations since they are available in digital form. In the context of shared annotations, notification mechanisms are applicable for advertising newly created and modified annotations. As will be discussed below, a crucial point is the underlying media format for the annotation system. Most annotation systems are designed for the Web, i.e. using HTML documents. However, there are also systems—often integrated into the proper application—for annotating PDF files as well as DOC and XML-based files, and so forth.

In summary, annotation systems are based on psycho-sociological and pedagogical learning theories. In terms of design, they are user-friendly offering search and notification functionalities and allow annotations to be shared or published for collaboration and cooperation.

Although there are conceptually as well as technologically highly sophisticated annotation systems, we have to ask why there is no prevalent annotation service today. Although most learning material is available in digital form, studies show that for annotation purposes people still prefer pen and paper [4].

In fact, this does not mean that people do not annotate in computer-based environments. People just use different kinds of tactics—*tactics* in the sense of the misuse of existing methods of one context in another context—to make annotations. People typically annotate using bookmarks, writing themselves email messages, posting in forums, and other similar tactics. However, why do not they use one of the existing annotation services or systems?

In this paper, we want to emphasize that annotation systems suffer from several fundamental problems. In Section 3, we argue that these problems are anomalies in terms of Thomas Kuhn's theory [5]. Hence, the contribution of this paper is as follows:

- We study annotation systems according to representative criteria.
- We identify the fundamental problems of existing annotation systems.
- We argue that these problems are anomalies and propose a shift of the annotation paradigm as a solution.
- We suggest the use of *artefacts* and *contexts*, a concept of a computer-oriented annotation paradigm.

In the subsequent section, we investigate annotation systems and services. In Section 3, we identify the fundamental problems of annotation systems. Section 4 introduces a computer-oriented annotation paradigm. Furthermore, it illustrates that the anomalies disappear using that paradigm. The status and future work followed by a conclusion finish the paper in Section 5 and Section 6, respectively.

## 2. Annotation Systems: A Study

In order to understand the deficiencies of the current annotation paradigm, it is essential to understand the workings of some aspects of today's annotation systems. This section compares different annotation systems and services based on selected characteristics as shown in Table 1. The list of annotation system is non-exhaustive. Our objective was to demonstrate that during the evolution of annotation systems aspects of those systems obeyed a consistent process, which persists even to this day.

Davis et al. [6] have developed CoNote, a Web-based system that allows collaborative writing of didactical material. The use of the CoNote system supports discussions after lesson time and exchanges between students.

Harger [7] developed and approved the idea of a digital book that, integrated with software for mathematical simulation, allows creation of personal annotations.

Sparrow [8] allows a collaborative editing of shared community Web pages. Annotations are considered distinct entities but do not exist independently of the document. There are three textual annotation types: list, structured page (outline-like), and fixed form.

Additional systems that are fully Web-based include ComMentor [9], CoNote [6], CritLink

**Table 1: Comparison of annotation systems and services**

| | Sparrow | ComMentor | CoNote | CritLink Mediator | Annozilla | Adobe Acrobat | WORD | Flickr |
|---|---|---|---|---|---|---|---|---|
| **Media format** | HTML | HTML | HTML | HTML | HTML | PDF | DOC, RTF | JPG |
| **Annotation appearance** | In-line | In-place | Anchor points | In-place | In-place | In-place) | In-place | In-place |
| **Classification of annotation** | List, outline, fixed form | SampleSet Tri-level Paper Comments | Author Reader | In-place In-line | Question, Explanation, Advice, Change, Example, seeAlso, Comment | Audio Text Drawing-(objects) | Reviewer | Tags, notes, comments |
| **Reply annotation** | Yes | No | No | Yes | No | Yes | No | No |
| **Searchable** | ? | ? | Yes | No | No | Yes | Yes | Yes |
| **Shareable** | ?/?/? | Private Group Public | Public | Public | Private Group | Public | Public | Public |
| **Handling of artefacts** | Repository | Repository | | Repository on server OR on client | Repository on server OR on client | Repository/ Attached on the source | Attached on the source | |
| **Notification mechanism** | ? | No | No | ? | ? | No | No | No |
| **Technology** | -- | CGI (Perl), PRDM | CGI (Perl) | CGI-Scripts | RDF, Xpointer | -- | -- | -- |

Mediator [10] and Annozilla [11]. In these systems, both the sources as well as annotations are HTML documents. This provides the possibility of linking to other pages. Adobe Acrobat's commenting service extends this functionality by directly allowing recorded audio comments within the PDF files [12].

CoNote was designed to provide an easy approach for coordinating collaborative work. In contrast, ComMentor annotations are stored separately from the source documents. The author of the source document defines the anchor points (place-of-annotation concept). There is just one type of annotations, but it is possible to differentiate annotations from the author and the readers.

Like most of the approaches, Annozilla is also open-source. However, it suffers from a missing annotation concept. Annotation types like questions, explanation, advice, example, etc., are available, but in terms of representation or features, they do not differ from each other.

CritLink was created to support critical discussions. Annotations can be added at any point in the document. It is also possible to reply to annotations this way creating an annotation structure dynamically. There are two types of annotations: place-of-annotation (in-line) and on the end of the document (on-page). Furthermore, CritLink suffers from inappropriate reply functionality to annotations. Other systems do not allow replying to annotations [11, 13]. In general, it is hard to keep hierarchical annotations and replies visually manageable. Adobe Acrobat is one such example [12].

Flickr [14] is a public Web-based photo management system. The integrated annotation service allows image comments. Information about the image is represented as a comment. Tags and notes specify a region of the image with meta-data.

Microsoft Word [15] can store comments from different reviewers in its DOC files. Comments are in-place and can be made via a split screen or a bubble mode. It is possible to search in the content of text, including the annotations, but impossible to search in the comments only. More sophisticated functionality like sharing of annotations, repositories, or reply notification mechanisms are not provided.

## 3. Anomalies in the Annotation Paradigm

As presented in the introduction we have asked why there are no prevalent annotation services today while sophisticated annotation systems have existed for some time now. Furthermore, even though most learning material is available in digital form, studies show that for the purpose of annotating, people still prefer to use pen and paper [4]. However, people make computer-based annotations using bookmarks, email messages, newsgroup messages, and alike. Why do not they use one of the existent annotation services or systems? In the subsequent sections, we tackle this question discussing four main problems with conventional annotation systems.

*Different services for one activity.* Annotation systems are commonly designed for a single media format, like PDF, DOC, or HTML (cf. Table 1). This means that learners that use learning material in different media formats have to use several annotation services in conjunction. Annotating is not a collateral effect of reading, but a significant cognitive activity that is not taken into account in the design of existing annotation systems.

*Reading and thinking occurs not only linearly, but also mostly laterally.* People open several documents to get informed. They are reading multiple, complementary sources and documents at the same time whereby the main focus can shift from one document to another applying lateral thinking [16]. They even open audio files to listen to interviews or videos of a presentation they could not attend. Consider a scenario where two documents contain contradictory information (cf. Fig. 1). How can a reader be enabled to track down his doubts in a natural way? Cognitive efforts spent for relating two or more documents, probably even of different types, cannot be captured at all and will be lost.

*Collaboration creates verbal and gestural annotation.* Many annotations have never been written down, because they have a verbal or gestural nature. We use this form during meetings or collaborations for instance. With technologies, collaboration can occur between people far away from each other. Sometimes they cannot meet at the same time and this particular form of annotating is impossible to grasp. Asynchronous communication technologies try to close this gap, but the common annotation systems do not support annotations that capture movements, pointing, gestures, and alike.

*Time-binding character of collaborations and chronological order of interactions.* To understand the intention of an annotation it is important to be aware of the context the reader has worked in. When did the learner open documents A and B? Did the learner read the whole document or just a paragraph? Did the learner apply more linear reading or lateral reading? Of which kind were the interactions? During synchronous collaboration, this knowledge is tacit, but exists and the participants understand each other. In an asynchronous environment, conventional annotation systems fail to be aware and keep track of these kinds of tacit information.

In our opinion, changing the concept or adjusting functionalities of existing annotation systems cannot resolve problems of such a fundamental character. Even a new annotation system is doomed to fail if it is based on the pen-and-paper annotation paradigm only. Our approach is to interpret these fundamental problems as anomalies. Anomalies are failures of the current paradigm as described by Thomas Kuhn [5]. In the literature, there is often the statement that there is no evolution regarding annotation systems. This is because the pen-and-paper annotation paradigm in computer-based environments has been stretched to its very limits. Consequently, a paradigm shift has to be provoked.

## 4. Shifting the Annotation Paradigm via Artefacts and Contexts

### 4.1. Artefact and Context

In order to overcome the anomalies identified in Section 3, we introduce the artefact-context annotation paradigm that is supposed to be adapted to computer-based annotations. In this paradigm, an annotation is augmented by additional semiotic functionalities enabling pointing and gesture-like annotations as well as additional attributes. Due to this augmentation, we call these sophisticated computer-based annotations artefacts. As will be discussed, artefacts go beyond the conventional idea of *annotations*.

As an annotation manifests itself on a media, an artefact also needs to do so. When annotating a book page, the media is the paper. We define the media of an artefact as its context. Artefacts are not independent pieces of data, but rather created and applied in a clearly defined context. Why does the context appear to be so important? When managing artefacts, their context has to be preserved, as the page of a book with an annotation has to be preserved.

Artefacts are searchable, just like annotations, but the immanent nature of artefacts reveals a higher level of complexity. Thus, artefacts preserve the interactions on the context chronologically. In order to facilitate that, the context consists of substantially more than just the information about the current digital documents. Rather, the context tracks data about running applications and their open documents. In order to track the history of interactions activation, instantiation, and idle time is of essential importance as well as the tracking of modifications on documents. Creating an artefact means to preserve a snapshot of the state of current context.

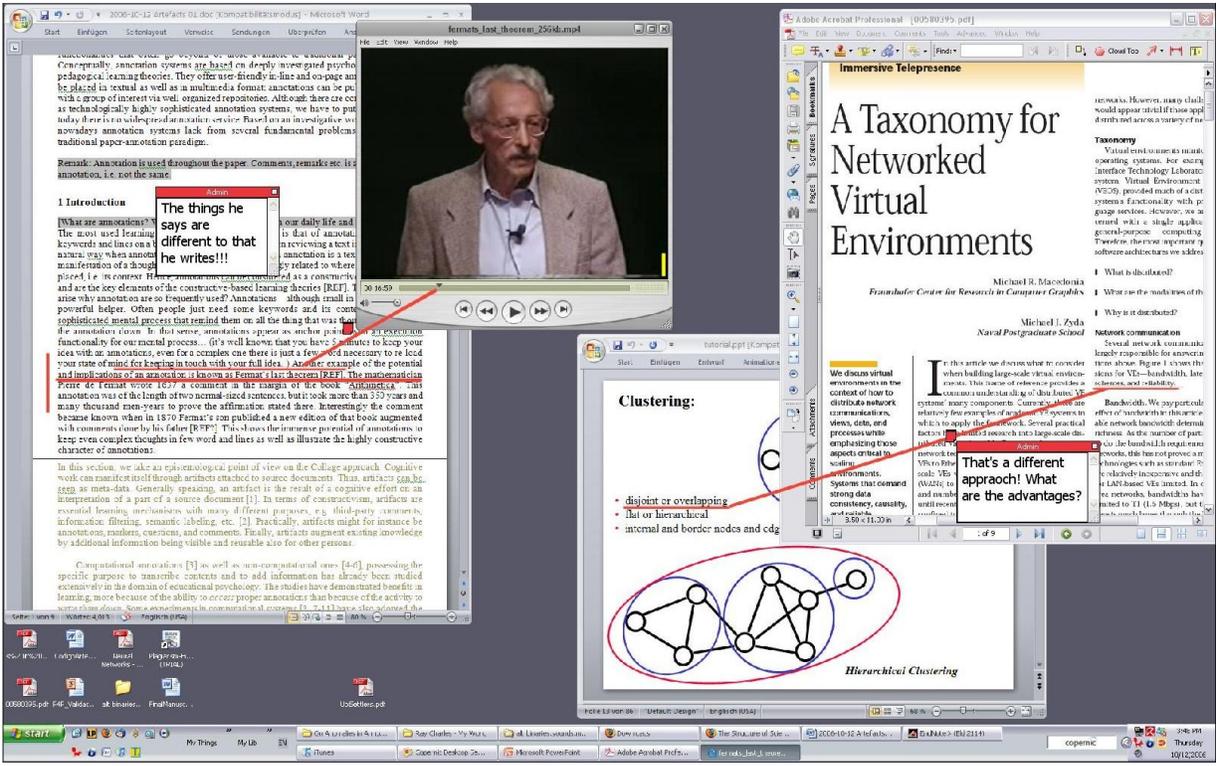

**Figure 1: Illustration of artefacts (red) and their contexts (documents, interactions, etc.)**

If the reader creates an artefact on a context where an artefact already exists, then—by definition of a context—the earlier artefact turns into a part of that context. Additionally, a context can be used as an artefact, since an artefact can be anything, even a context. These two circumstances—that an artefact can be turned into a part of a context and together with a context can be used as artefact—reveals a duality between artefact and context, being in fact the same entity. It depends on the usage as this entity is understood, either in the view of an artefact or as context. If this duality really exists, we can take advantage of this fact by using the same technology for implementing both artefacts and contexts. Aside from that, it also would mean that our theory is closed and complete, no additional elements or entities need to be considered.

### 4.2. Dealing with Anomalies

In Section 3, we revealed anomalies of the pen-and-paper annotation paradigm that appear when being applied to the computer-based environment. The previous section introduces the artefact-context annotation system that is supposed to be more suitable for the purpose of implementing annotation systems. Subsequently, we show that the annotation paradigm introduced overcomes those anomalies.

*One service for all annotating activities.* Through to the introduction of the context concept, the media of an annotation (pen-and-paper paradigm) was incrementally augmented beyond the limits of one document or one application. A context can be the entire current working space, e.g. everything that resides and happens on the desktop. It can be considered as a kind of independent overlay. Consequently, the application of an artefact is a comprehensive coverage of any event on that overlay. Thus, the user can realize any annotation activities by one annotation service—independently from the document format or application.

*Lateral annotations.* As described before, an annotation service that implements the artefact-context annotation paradigm is able to realize annotations equally to different media. A logical consequence of that circumstance is to make use of *lateral annotations* that link different sources, i.e. documents from several applications, in a sophisticated way by using a context as media. When a student opens a document such as slides from a lecture and a PDF file the student may become aware of some misconceptions. The student may decide to open the lecture script and watch a lecture covering that topic. As discovering some inconsistencies, the student creates an artefact linking those two documents. As seen in Fig. 1, one can annotate across the broad spectrum of publishing programs and documents available on the desktop in form of artefacts. One inherent benefit of the artefact appears essentially by keeping its context. That is the reason why artefacts go beyond the idea of conventional annotations.

*Capturing verbal and gestural annotations.* Artefacts cannot be realized by the common hyperlink techniques, because the context would be mostly lost. The importance of context dictates that all relevant documents and states for an artefact be preserved during the artefact creation process. In addition, since the context is able to keep traces of interactions, gestural or "tacit" annotations are potentially feasible.

*Immanent time-binding character.* In a context interactions are fully tracked in a history and eventually captured by the application of an artefact, the time-binding character is immanent to systems that follow the artefact-context annotation paradigm. A further difference of the conventional annotation and the artefact-and-context annotation paradigm is the time-binding character. The time-binding aspect is inherent to the computer, but not to a book.

### 4.3. Discussion

Maybe the annotation principle has progressed in a healthy way, but the importance of the context was underestimated. Therefore, our approach focuses on artefact and context adhering to the same rights. Including the context as a first class citizen of the theory, a time-binding character may be more suitable for application in computer-based environments.

Practically speaking, artefacts and contexts are overcoming the lack of interaction mechanisms in current computer supported collaborative learning systems. Since [17] we pointed out that interaction mechanisms for didactic material are insufficient for supporting collaborative work between students, nothing fundamental has changed until the present time. As mentioned already, most annotation systems provide annotation mechanisms for just one document format, e.g. for HTML, PDF, DOC, RTF. More severe is the limitation to annotate a single document. Cognitive efforts spent on relating two or more documents, probably of different types, cannot be captured at all by applying the conventional pen-and-paper annotation paradigm.

Annotation systems following the artefact-context annotation paradigm explicitly allow introducing inter-document and inter-application annotations (lateral annotations) as illustrated in Fig. 1. This way, interrelating two or more source documents, even if they are of different media types, becomes possible as shown in the previous sub-section. Artefacts allow annotations to refer to a Web page and a video clip, for example. This is far beyond just employing hyperlinks between two Web pages. Artefacts and contexts are first class citizens of the system, having their own individual representation, implementable by using the same abstractions of their supposed duality.

## 5. Current Status and Future Work

The system we are introducing is called *COLLAGE*—Collaborative Artefact Gathering Environment. COLLAGE follows the constructivism theory and provides annotations for multiple source types, i.e. realizing artefacts together with their enclosing contexts.

COLLAGE is designed to create new potentials for capturing cognitive efforts and publishing them within closed groups as well as across group boundaries. Students will be enabled to review the definitions, explanations and comments of fellow students on any given subject. Attaching information presenting the viewpoints of well-known specialists in the field can also be achieved more easily. In particular, joining efforts within groups can stimulate further information gathering. Through the system, questions can be asked and answered anytime, anywhere. Students can also spontaneously join local groups in the neighborhood. In summary, targeted at mobile platforms, the way in which COLLAGE permits interactions makes it well suited to the field of collaborative learning and annotation environments.

For the time being, building blocks of COLLAGE have been developed prototypically. At the level of artefacts, we currently cover, among others, all types of OpenOffice documents. A special kind of link can be used to point to particular parts of an OpenOffice document. Such links can be generated from the current viewing context or viewport. Navigating to the target of such links is possible as well, i.e. opening the underlying document and navigating to the location pointed to by the link. Similar actions are possible with the Firefox browser. Support for PDF documents is currently under construction. This way, basic support for capturing contexts and managing artefacts exists. However, more sophisticated functionalities, especially those for capturing and re-playing the timeline of contexts still needs to be developed.

Building, managing, and exploiting interest groups is investigated in the HyMN application [18]. Hereby, the focus is on the distribution of multimedia content which might encompass artefacts and contexts.

## 6. Conclusion

The same way as the word-processing paradigm has substituted the typewriter paradigm, the pen-and-paper annotation paradigm is shifting towards the artefact-and-context paradigm introduced in this paper.

Despite its ubiquity and strengths, the paper-and-pen approach is limited by today's media standards. Paper is well-suited for communicating static images, but electronic media can incorporate motion and sound, and can thus communicate certain information more clearly. As we show, the pure pen-and-paper paradigm permanently creates anomalies when being applied on computer-based scenarios.

In order to avoid these anomalies, we argue that it is necessary to shift the annotation paradigm. Our approach is to introduce the artefact-context annotation paradigm that is supposed to be more suitable for the purpose of implementing annotation systems. We illustrated that the artefact-context annotation paradigm can overcome those anomalies.

The artefact-context annotation paradigm as introduced throughout this paper certainly serves as a starting point only. It has to be refined and verified by being used for implementing an annotation system.

## 7. References

[1] D. Jonassen, "Constructivism and Computer-Mediated Communication in Distance Education," in *Constructivism and the Technology of Instruction: A Conversation*. vol. 9, 1995, pp. 7-26.

[2] T. C. Ahern, "Using online annotation software to provide timely feedback in an introductory programming course," in *35th IEEE Frontiers in Education Conference*, 2005.

[3] P. Ribenboim, *13 Lectures on Fermat's Last Theorem*: Springer, 1995.

[4] D. Bargeron and T. Moscovich, "Techniques for on-screen shapes, text and handwriting: Reflowing digital ink annotations," in *SIGCHI conference on Human Factors in computing systems*, 2003.

[5] T. S. Kuhn, *The Structure of Scientific Revolutions*, 3rd ed.: University of Chicago Press, 1996.

[6] J. R. Davis and D. P. Huttenlocher, "Shared Annotations for Cooperative Learning," 1994.

[7] R. O. Harger, "Teaching in a Computer Classroom with a Hyperlinked, Interactive Book," *IEEE Transactions on Education,* vol. 93, pp. 327-335, 1996.

[8] B.-W. Chang, "In-place editing of Web pages: Sparrow community-shared documents " in *Computer Networks and ISDN Systems*, 1998, pp. 489-498.

[9] M. Röscheisen, C. Mogensen, and T. Winograd, "Beyond Browseing: Share Comments, SOAPs, Trails, and On-line Communites," 1995.

[10] R. M. Heck, S. M. Luebke, and C. H. Obermark, "A Survey of Web Annotations," 1999.

[11] M.-R. Koivunen, "Annotea and Semantic Web Supported Collaboration ", 2004.

[12] M. Ron and P. Kaari, "From tools to tasks: discoverability and Adobe Acrobat 6.0," in *CHI '04 extended abstracts on Human factors in computing systems*, 2004.

[13] G. W. Patricia, M. N. Christine, and B. Barbara, "Effects of interfaces for annotation on communication in a collaborative task," in *Proceedings of the SIGCHI conference on Human factors in computing systems*, 1998.

[14] W. Jill, "Feral hypertext: when hypertext literature escapes control," in *Proceedings of the sixteenth ACM conference on Hypertext and hypermedia* Salzburg, 2005.

[15] B. S. Gary, J. C. Thomas, and E. V. Misty, *Microsoft Office Word 2003: Complete Concepts and Techniques, CourseCard Edition*: Course Technology Press, 2005.

[16] S. Waks, "Lateral Thinking and Technology Education," *Journal of Science Education and Technology,* vol. 6, pp. 245-255, 1997.

[17] R. d. Oliveira, *Informática Educativa: Dos planos e discursos à sala-de-aula*: Editora Papirus, 1997.

[18] A. Andronache, M. R. Brust, and S. Rothkugel, "Multimedia Content Distribution in Hybrid Wireless using Weighted Clustering," in *2nd ACM Workshop on Wireless Multimedia Networking and Performance Modeling*, 2006.